\documentstyle[twoside,cosmion,epsf]{article}

\newcommand {\Mpc}   {\mbox{$h^{-1}$ Mpc \,}}

\newcommand{\mincir}{\raise -2.truept\hbox{\rlap{\hbox{$\sim$}}\raise5.truept
\hbox{$?$}\ }}
\newcommand{\gr}{\kern 2pt\hbox{}^\circ{\kern -2pt K}} %  ====? GRADI KELVIN
\newcommand{\magcir}{\raise -2.truept\hbox{\rlap{\hbox{$\sim$}}\raise5.truept
\hbox{$?$}\ }}

\newcommand{\Om}{\Omega}

\newcommand{\be}{\begin{equation}}
\newcommand{\ee}{\end{equation}}
\newcommand{\bea}{\begin{eqnarray}}
\newcommand{\eea}{\end{eqnarray}}

\newcommand{\si}{\sigma}

\newcommand{\etal}{{et al.}}

\heads{Determination of cosmological parameters} 
{Large scale structure of the Universe \ \ {\rm B. Novosyadlyj et al.}}

\begin{document}

\Arthead{1}{1}% Do not change this line

\Title{Determination of cosmological  parameters  from
 	large scale structure observations}
{B.~Novosyadlyj$^1$, R.~Durrer$^2$,
S.~Gottl\"ober$^3$, V.N.~Lukash$^4$, S.~Apunevych$^1$}
{$^1$Astronomical Observatory of L'viv State University, Kyryla and
Mephodia str.8, 290005, L'viv, Ukraine \\
$^2$Department de Physique Th\'eorique, Universit\'e de Gen\`eve,
Quai Ernest Ansermet 24, CH-1211 Gen\`eve 4, Switzerland \\
$^3$ Astrophysikalisches Institut Potsdam, An der Sternwarte 16,
D-14482 Potsdam, Germany \\
$^4$Astro Space Center of Lebedev Physical Institute of RAS,
Profsoyuznaya 84/32, 117810 Moscow, Russia}

%
%%%%%%%%%%%%%%%%%%%%%%%%%%%%%%%%%%%%%%%%%%%%%%%%%%%%%%%%%%%%%%%%%%%%%%%%%%

\Abstract{%

The possibility of determining cosmological parameters on the basis of a wide set
of observational data including the Abell-ACO cluster power spectrum and mass
function, peculiar velocities of galaxies, the distribution of Ly-$\alpha$ clouds
and CMB temperature fluctuations is analyzed.  Using a $\chi^2$ minimization method,
assuming $\Omega_{\Lambda}+\Omega_m =1 $ and no contribution from gravity
waves, we found that a tilted $\Lambda$MDM model with one sort
of massive neutrinos and the parameters $n=1.12\pm 0.10$, $\Omega_m=0.41\pm 0.11$
($\Omega_{\Lambda}=0.59\pm0.11$), $\Omega_{cdm}=0.31\pm 0.15$,
$\Omega_{\nu}=0.059\pm 0.028$, $\Omega_b=0.039\pm 0.014$ and $h=0.70\pm 0.12$
(standard errors) matches  observational data best. 
$\Omega_{\nu}$ is higher for more species of massive neutrinos, $\sim 0.1$
for two  and $\sim 0.13$ for three species. $\Omega_m$ raises by $\sim 0.08$
and $\sim 0.15$ respectively.
The 1$\sigma$ (68.3\%) confidence limits on each cosmological parameter, which are
obtained by marginalizing over the other parameters, are $0.82\le n\le1.39$,
$0.19\le\Omega_m\le 1$ ($0\le\Omega_{\Lambda}\le 0.81$),
$0\le\Omega_{\nu}\le 0.17$, $0.021\le \Omega_b\le 0.13$ and
$0.38\le h\le 0.85$.
Varying only a subset of parameters and fixing the others
shows also that the observational data set used here rules out
pure  CDM models with $h\ge 0.5$, scale invariant primordial power
spectrum, zero cosmological constant and spatial curvature at a very
high confidence level, $>99.99\%$.
The corresponding class of MDM models are ruled out at $\sim 95\%$ C.L.
It is notable also that this data set 
determines the amplitude of scalar fluctuations approximately at the
same level as COBE four-year data. It indicates  that a possible
tensor component in the COBE data cannot be very substantial.  }

\section{Introduction}

The last years of the past century are marked by huge efforts of the
community of astronomers, physicists and astrophysicists devoted to determine
the most fundamental parameters of our Universe, the cosmological parameters.
The most important among them  are the
mass densities of baryons $\Omega_b$ (in units of the critical density) and   
of cold dark matter $\Omega_{cdm}$, the neutrino rest masses  $m_{\nu}$
and their total density $\Omega_{\nu}$, the value of cosmological
term $\Lambda$ (or $\Omega_{\Lambda}$), the Hubble constant $H_0$,
the spatial  curvature parameter $\Omega_k$ and the
slopes $n$ and amplitudes $A$ of the primordial power spectra of 
scalar and tensor fluctuations.

The primordial ratio of the number of deuterium to hydrogen nuclei (D/H) 
created in Big Bang nucleosynthesis is the most sensitive measure of the 
cosmological density of baryons $\Omega_b$. Quasar absorption systems give
definite measurements of the primordial deuterium and the most accurate value 
of baryon density obtained recently in this way is $\Omega_bh^2=0.019\pm0.0024$
\cite{bur99}.

The measurements of the neutrino rest mass is not so certain, unfortunately.
Up-to-day we have only some indications for the range where it may be found.
The oscillations of solar and atmospheric neutrinos registered by the 
SuperKamiokande experiment show that the difference of rest masses between 
$\tau -$ and $\mu$-neutrinos is $0.02<\Delta m_{\tau \mu} < 0.08eV$
\cite{fu98,pr98}. This also provides a lower limit for the neutrino
mass,   $m_{\nu}\ge
|\Delta m|$ and does not exclude models with  cosmologically
significant values $\sim 1-20eV$. Therefore, at least two  species of
neutrinos can have approximately equal masses in this range.  Some
versions of elementary particle theories predict  $m_{\nu _e}\approx
m_{\nu _\tau}\approx 2.5eV$ and  $m_{\nu _{\mu}}\approx m_{\nu _s}\sim
10^{-5}eV$, where  ${\nu _e}$, ${\nu _\tau}$, ${\nu _{\mu}}$ and ${\nu
_s}$ denote the  electron, $\tau -$, $\mu -$  and sterile neutrinos
accordingly (e.g. \cite{dol95}).  The strongest 
upper limit for the neutrino mass comes from the observed large scale 
structure of our Universe: $\sum_{i} m_{\nu_i}/94{\rm eV}\le 0.3h^2$.
Since observations give for the Hubble parameter an upper limit of
$h=0.8$ one gets $\sum_{i} m_{\nu_i}\le 18eV$. It is interesting to note that
this  upper limit coincides roughly with the upper limit for the electron
neutrino mass obtained from the supernova explosion 
SN1987A and tritium $\beta$-decay experiments.
 
Important conclusions about measurements of matter density
$\Omega_m$ ($\equiv \Omega_b+\Omega_{cdm}+\Omega_{\nu}$)
come from the Supernova Cosmology Project and the High-z Supernova Search.
In particular, the relation of observed brightness vs. redshift for SNeIa
shows that distant supernovae are fainter than expected for a
decelerating Universe, and, thus, more distant. This can be
interpreted as an accelerated  expansion
rate, or $\Omega_{\Lambda}>0$. The best-fit value is
$\Omega_{\Lambda}={4\over 3}\Omega_m + {1\over 3}\pm 0.1$
(1$\sigma$ error) and  $\Omega_m=1$ models are ruled out
at the 8$\sigma$ level \cite{per98}. For a flat Universe
$\Omega_m + \Omega_{\Lambda}=1$ ($\Omega_k=0$) the best-fit
values are $\Omega_m=0.25\pm 0.1$ and $\Omega_{\Lambda}=0.75\pm 0.1$
\cite{per98,rie98,bah99}.

An upper limit of $\Omega_{\Lambda}<0.7$ (95\% C.L.) follows from
gravitational lensing statistics\cite{bar98,fal98}, just
 consistent with distant supernovas results.

Strong evidence against an open Universe can be derived from recent
measurements of the position of the  first acoustic peak in
the cosmic microwave background (CMB) power spectrum by
the Boomerang experiment\cite{mau99}. The $1\sigma$ range for the curvature
parameter derived from this experiment is 
$-0.25\le \Omega_k\le 0.15$ \cite{mel99}
and the mean value is close to the flat Universe, $\Omega_k\approx 0$.

Currently there are a few completely independent and broad
routes to the determination of the Hubble constant $H_0$. The
direct experiments   can be divided into three
groups: the gravitational lens time delay methods,
the Sunyaev-Zel'dovich method for clusters and extra-galactic
distance measurements. Almost all observations yield values of $H_0$ 
in the  range 50-80 km/sec/Mpc.

Other independent methods for the determination of cosmological
parameters are based on  large scale structure (LSS)
observations. Their advantage is that all parameters mentioned
above can be determined together because the form and amplitude
of the power spectrum of density fluctuations are rather sensitive
to all of them. Their disadvantage is that they are model dependent.
This approach has been carried out in several
papers (e.g. \cite{atr97,lin97,Teg99,bri99,nov99,phill} and references therein)
and it is also the goal of this paper. The papers on this subject differ by the
number of parameters and the set of observational data included into the
analysis. In this paper a total of 23 measurements from sub-galaxy
scales (Ly-$\alpha$ clouds) over cluster scales up to horizon scale
(CMB quadrupole) is used to determine eight cosmological parameters,
namely the tilt of the primordial spectrum $n$, the densities of
cold dark matter $\Omega_{cdm}$, hot dark matter $\Omega_{\nu}$,
baryons $\Omega_b$ and cosmological constant $\Omega_{\Lambda}$,
the number of massive neutrino species $N_{\nu}$, the Hubble parameter
$H_0$ and, in addition, the bias parameter $b_{cl}$ for rich clusters 
of galaxies.
 We restrict ourselves to the analysis of spatially flat cosmological
models with $\Omega_{\Lambda}+\Omega_m=1$ ($\Omega_k=0$) and to an inflationary
scenario without tensor mode. We also neglect the effect of a possible
early reionization which could reduce the amplitude of the first
acoustic peak in the CMB anisotropy spectrum.

In comparison to the companion paper\cite{nov00} the influence
of the uncertainties in the normalization of the scalar mode amplitude
caused by experimental errors on determination of cosmological parameters
is also taken into account  here.

\section{The experimental data set and our methods}

We use the power spectrum of Abell-ACO clusters 
\cite{ein97,ret97}, measured in the range
$0.03\le k\le 0.2h/$Mpc, as observational input. 
Its amplitude and slope at lower and larger scales
are quite sensitive to baryon content $\Omega_{b}$,
Hubble constant $h$, neutrino mass $m_{\nu}$ and number of species of
massive neutrinos $N_{\nu}$ \cite{nov99}.
The total number of Abell-ACO data points with their errors used for minimization
is 13, but not all of these points can be considered as independent 
measurements. Since we can accurately fit the power spectrum by an analytic 
expression depending on three parameters only (the amplitude at large 
scales, the slope at small scales and the scale of the bend); we
assign to the power spectrum 3 effective degrees of freedom.

The second observational data set which we use are 
the position and amplitude of the first acoustic peak derived 
from the data on the angular power spectrum of CMB temperature fluctuations.
To determine the position and amplitude of the
first acoustic peak we use a 6-th order polynomial fit to
the data set on CMB
temperature anisotropy, accumulated in Table 2 of our accompanied paper 
\cite{nov00}, 51 data points in total. 
The amplitude $A_p$ and position $\ell_p$ of first acoustic peak
determined from this fit are $79.6\pm 16.5\mu \rm K$ and $253\pm 70$
correspondingly.
The statistical errors are estimated by edges of the $\chi^2$-hyper-surface 
in the space of polynomial coefficients which corresponds to 
68.3\% ($1\sigma$) probability level under the assumption of
Gaussian statistics. 
Also the mean weighted bandwidth of each experiment around 
$\ell_p$ is added to obtain total $\Delta \ell_{p}$.

A constraint on the amplitude of the matter density fluctuation power spectrum at
cluster scale can be derived from the cluster mass and X-ray temperature
functions. It is usually formulated in terms of the density
fluctuation in a top-hat sphere of 8\Mpc radius, $\sigma_{8}$, which
can be easily calculated for the given initial power spectrum.
According to the recent optical determination of the mass function of nearby 
galaxy clusters \cite{gir98} and taking into account the results from other authors
(for references see \cite{borg99}) we use the value  
$\tilde \sigma_{8}\tilde\Omega_m^{0.46-0.09\Omega_m}=0.60\pm 0.08$.
>From the existence of three most massive clusters of galaxies observed
at $z>0.5$ a further constraint has been established by Bahcall \& Fan
\cite{bah98}: $\tilde \sigma_8\tilde\Omega_m^{\alpha}=0.8\pm 0.1\;,$
where $\alpha =0.24$ if $\Omega_{\Lambda}=0$ and $\alpha =0.29$ if
$\Omega_{\Lambda}>0$ with $\Omega_{\Lambda}+\Omega_m=1.$ 

A constraint on the amplitude of the linear power spectrum of
density fluctuations in our vicinity comes from the study of galaxy
bulk flow, the mean peculiar velocity of galaxies in sphere of radius 
$50h^{-1}$Mpc around our
position. We use the data given by Kollat \& Dekel \cite{kol97}, $\tilde V_{50}=(375\pm 85)$ km/s.

A further essential constraint on the linear power spectrum of matter
clustering at galactic and sub-galactic scales 
$k\sim (2-40)h/$Mpc can be obtained
from the Ly-$\alpha$ forest of absorption lines seen in quasar spectra
\cite{gn98,cr98}.  Assuming that the
Ly-$\alpha$ forest is formed by discrete clouds of a physical extent
near Jeans scale in the reionized inter-galactic medium at $z\sim 2-4$,
Gnedin \cite{gn98}  has obtained a constraint on the value of the
r.m.s. linear density fluctuations $1.6<\tilde \sigma_{F}(z=3)<2.6$
(95\% C.L.) at Jeans scale for $z=3$ equal
 to $k_{F}\approx 38\Omega_m^{1/2}h/$Mpc \cite{gn99}.

The procedure to
recover the linear power spectrum from the Ly-$\alpha$ forest has been
elaborated by Croft et al.\cite{cr98}. Analyzing the absorption lines in a sample
of 19 QSO spectra, they have obtained the following 95\% C.L. constraint on the
amplitude and slope of the linear power spectrum at $z=2.5$ and
$k_{p}=1.5\Omega_m^{1/2}h/$Mpc
\begin{equation}
\tilde \Delta_{\rho}^2(k_p)\equiv k_p^3P(k_p)/2\pi^2=0.57\pm 0.26,
\label{Deltarho}
\end{equation}
\begin{equation}
\tilde n_p\equiv {\Delta \log\;P(k)\over \Delta \log\;k}\mid_{k_p}=-2.25\pm 0.1.
\label{n_p}
\end{equation}

In addition to the power spectrum measurements we use
the constraints on the value of Hubble constant $\tilde h=0.65\pm
0.15$ which is a compromise between measurements made by two groups:
\cite{Tamm99} and \cite{mad98}. We also employ the 
nucleosynthesis
constraints on the baryon density of $\tilde{\Omega_bh^2} =
0.019\pm 0.0024$ (95\% C.L.) \cite{bur99}.

In order to find the best fit model we must evaluate the 
above mentioned quantities for a given cosmological model. 

To  this end we use the accurate analytic
approximations of the MDM transfer function $T(k;z)$ depending on
the parameters $\Omega_m$, $\Omega_b$, $\Omega_{\nu}$, $N_{\nu}$, 
$h$ by Eisenstein \& Hu \cite{eh3}.

The linear power spectrum of matter density fluctuations is
\begin{equation}
P(k;z)=Ak^nT^2(k;z)D_1^2(z)/D_1^2(0),\label{pkz}
\end{equation}
where $A$ is the normalization constant and 
$D_1(z)$ is the growth factor,
useful analytical approximation for which has been given by Carrol et al.\cite{car92}.

We normalize the spectra using the 4-year COBE  data  which can be
expressed by the value of the density perturbation at the horizon crossing
scale, $\delta_h$ \cite{lid96,bun97}. 
The normalization constant is related to $\delta_h$ by 
\be
A=2\pi^{2}\delta_{h}^{2}(3000/h)^{3+n}
\;{\rm Mpc}^{3+n}.\label{anorm}
\ee

The Abell-ACO power spectrum is given by the matter power
spectrum at $z=0$ multiplied by a linear and scale independent
cluster biasing parameter $b_{cl}$, which we include as a free parameter
\begin{equation}
P_{A+ACO}(k)=b_{cl}^2 P(k;0).\label{bias}
\end{equation}

For a given set of parameters $n$, $\Omega_m$, $\Omega_b$, $h$,
$\Omega_{\nu}$, $N_{\nu}$ and $b_{cl}$ the theoretical value of
$P_{A+ACO}(k_j)$ can now be calculated for each observed scale $k_j$.
Let's denote these values by $y_j$ ($j=1,...,13$).

The dependence of position and amplitude of the first acoustic
peak in the CMB power spectrum on cosmological
parameters has been investigated using the public code CMBfast by Seljak \&
Zaldarriaga\cite{sz96}.  As expected, these characteristics are
independent on the hot dark matter content.
We determine the values $\ell_p$ and $A_p$ for given parameters 
($n$, $h$, $\Omega_b$ and $\Omega_{\Lambda}$) on a
4-dimensional grid for parameter values in between the grid points we
determine  $\ell_p$ and $A_p$  by linear  interpolation.
We denote $\ell_p$ and $A_p$ by $y_{14}$ and $y_{15}$ respectively.

The theoretical values of the other experimental constraints are
obtained as follows: The density fluctuation $\sigma_8$ is
calculated according to
\begin{equation}
\sigma_{8}^{2}={1\over2\pi^{2}}\int_{0}^{\infty}k^{2}P(k;0)W^{2}(8{\rm Mpc}\;k/h)dk,
\end{equation}
with $P(k;z)$ from Eq.~(\ref{pkz}). We
set $y_{16}=\sigma_{8}\Omega_m^{0.46-0.09\Omega_m}$ and $y_{17} =
\sigma_{8}\Omega^{\alpha}$, where $\alpha =0.24$ for
$\Omega_{\Lambda}=0$ and $\alpha =0.29$ for $\Omega_{\Lambda}>0$,
respectively.

The r.m.s. peculiar velocity of
galaxies in a sphere of radius $R=50h^{-1}$Mpc is
\begin{equation}
V^{2}_{50}={1\over 2\pi^{2}}
\int_{0}^{\infty}
k^2P^{(v)}(k)e^{-k^{2}R_{f}^{2}}W^{2}(50{\rm Mpc}~k/h)dk,  \label{V50th}
\end{equation}
where $P^{(v)}(k)$ is the density-weighted power spectrum for the 
velocity field \cite{eh3}, 
$W(50{\rm Mpc}~k/h)$ is a top-hat window function, and 
$R_{f}=12h^{-1}$Mpc is the radius of a Gaussian filter used for smoothing
of the raw data.
For the scales considered  $P^{(v)}(k)\approx (\Omega^{0.6}H_0)^2P(k;0)/k^2$.
We denote the r.m.s. peculiar velocity by $y_{18}$.

The value of the r.m.s. linear density perturbation from the formation of 
Ly-$\alpha$ clouds at redshift $z$ and scale $k_{F}$ is given by
\begin{equation}
\sigma_{F}^{2}(z)={1\over
2\pi^{2}}\int_{0}^{\infty}k^{2}P(k;z)e^{(-k/k_F)^2}dk \label{siF}.
\end{equation}
We set $\sigma_{F}^{2}(z=3) y_{19}$. 

The value of
$\Delta_{\rho}^2(k_p,z)$ and the slope $n(z)$ are obtained from the
linear power spectrum $P(k;z)$ by 
Eq.~(\ref{Deltarho}) and Eq.~(\ref{n_p}) at 
$z=2.5$ and $k_{p}=0.008H(z)/(1+z)({\rm km/s})^{-1}$, and are denoted by
 $y_{20}$ and  $y_{21}$ accordingly.

For all tests except Gnedin's Ly-$\alpha$ test
we use the density weighted transfer function $T_{cb\nu}(k,z)$
from \cite{eh3}. For $\sigma_F$ the function $T_{cb}(k,z)$ is used
according to the prescription given by Gnedin \cite{gn98}. Note, however, that
even in the model with maximal $\Omega_{\nu}$ ($\sim0.2$) the difference
between $T_{cb}(k,z)$ and $T_{cb\nu}(k,z)$ is less than 
$ 12\%$ for  $k\le k_p$.

Finally, the values of $\Omega_bh^2$
and $h$ are denoted by $y_{22}$ and $y_{23}$ respectively.

Under the assumption that the errors on the data points are Gaussian,
the deviations of the
theoretical values from their observational counterparts can be
characterized  by $\chi^2$:
\begin{equation}
\chi^{2}=\sum_{j=1}^{23}\left({\tilde y_j-y_j \over \Delta \tilde y_j}
\right)^2,     \label{chi2}
\end{equation}
where $\tilde y_j$ and $\Delta \tilde y_j$ are the experimental data
and their dispersions, respectively. The set of parameters $n$,
$\Omega_m$, $\Omega_b$, $h$, $\Omega_{\nu}$, $N_{\nu}$
and $b_{cl}$  are then determined by minimizing $\chi^2$ using the
Levenberg-Marquardt method \cite{nr92}. The derivatives of the predicted
values w.r.t the search parameters required by this method are
calculated numerically using a relative step size of $10^{-5}$.

This method has been tested and has proven to be reliable,
independent on the initial values of parameters and it has 
good convergence.\\[2mm]

\section{Results}

The determination of the parameters $n$, $\Omega_m$, $\Omega_b$, $h$,
$\Omega_{\nu}$, $N_{\nu}$ and $b_{cl}$ by the Levenberg-Marquardt $\chi^2$
minimization method is realized in the following way: we vary the
set of parameters $n$, $\Omega_m$, $\Omega_b$, $h$,
$\Omega_{\nu}$  and $b_{cl}$ and find the minimum of
$\chi^2$, using all observational data described in previous section.
 Since the $N_{\nu}$ is discrete, we repeat the
procedure three times for $N_{\nu}$=1, 2, and 3.  The lowest of the
three minima is the minimum of $\chi^2$ for the
complete set of free parameters. 

\begin{table*}[th]
\caption{Cosmological parameters determined for the tilted
$\Lambda$MDM model with one, two and three species of massive
neutrinos.}
\begin{center}
\def\onerule{\noalign{\medskip\hrule\medskip}}
\medskip
\begin{tabular}{cccccccc}
\hline
&&&&&&&\\
$N_{\nu}$     & $\chi^2_{min}$  &$n$  & $\Omega_m$&$\Omega_{\nu}$& $\Omega_b$ & $h$    & $b_{cl}$ \\ [4pt]
\hline
&&&&&&&\\
1  & 4.64&1.12$\pm$0.09&0.41$\pm$0.11&0.059$\pm$0.028&0.039$\pm$0.014&0.70$\pm$0.12&2.23$\pm$0.33\\
2  & 4.82&1.13$\pm$0.10&0.49$\pm$0.13&0.103$\pm$0.042&0.039$\pm$0.014&0.70$\pm$0.13&2.33$\pm$0.36\\
3  & 5.09&1.13$\pm$0.10&0.56$\pm$0.14&0.132$\pm$0.053&0.040$\pm$0.015&0.69$\pm$0.13&2.45$\pm$0.37\\ [4pt]
\hline
\end{tabular}
\end{center}
\end{table*}

\begin{figure*}[tp]
\epsfxsize=16truecm
\epsfbox{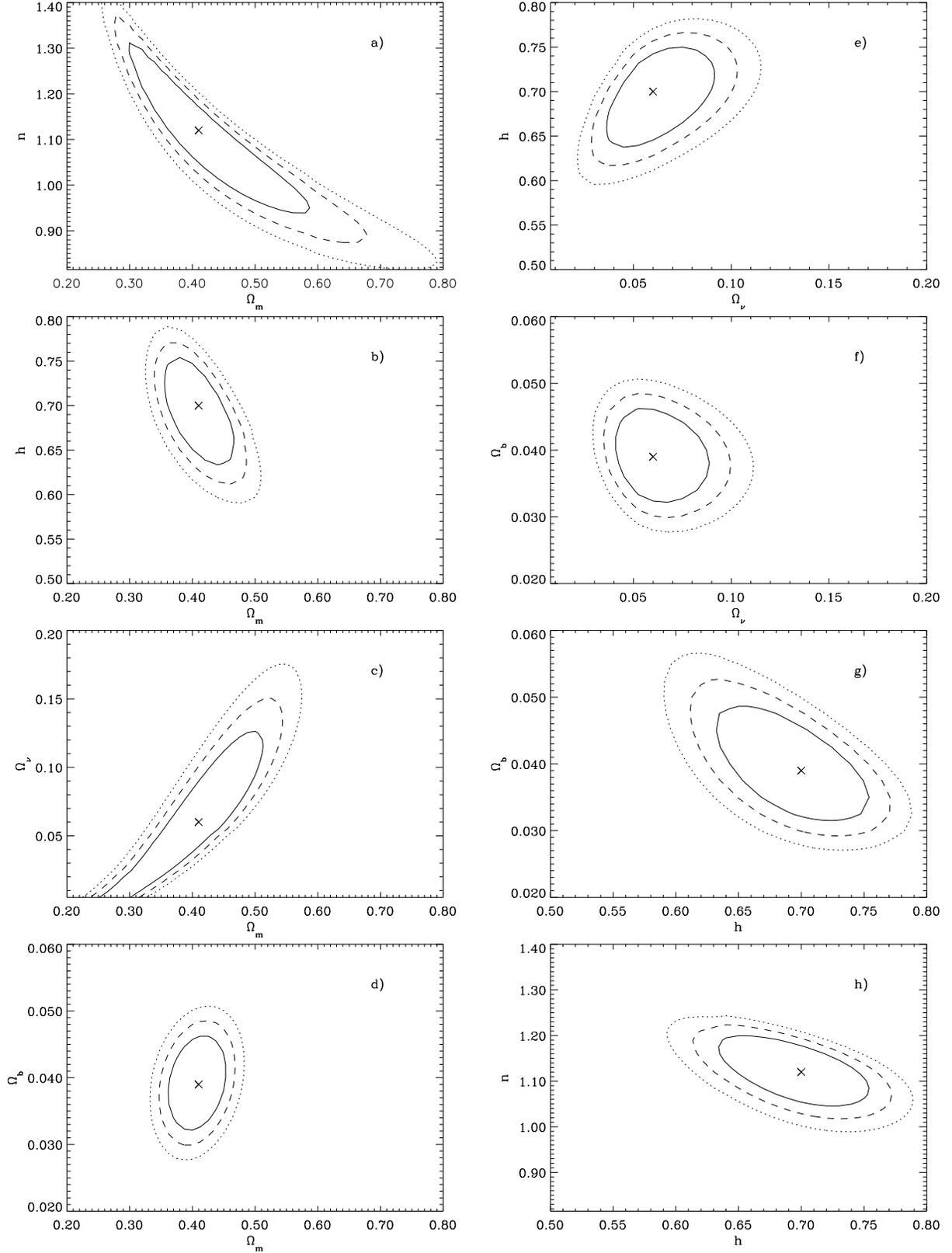}
\caption{Likelihood contours (solid line - 68.3\%, dashed - 95.4\%, dotted
- 99.73\%) of the tilted $\Lambda$MDM model with $N_{\nu}=1$ and
parameters from Table 1 ($N_{\nu}=1$) in the different planes of
$n-\Omega_m-\Omega_{\nu}-\Omega_b-h$ space.  The parameters
not shown in a given diagram are set to their best fit value.}
\label{Lcm1}
\end{figure*}

The results are presented in the Table 1. The errors in the determined
parameters are the square roots of diagonal elements of the covariance
matrix of standard errors. More information about the
accuracy of the determination of parameters and their sensitivity to
the data used can be obtained from the contours of confidence
levels presented in Fig.~\ref{Lcm1} for the tilted $\Lambda$MDM model with
parameters from Table 1 (case $N_{\nu}=1$).   These
contours show the confidence regions which contain 68.3\% (solid line),
95.4\% (dashed line) and 99.73\% (dotted line) of the total probability
distribution in the two dimensional sections of the six-dimensional
parameter space, if the probability distribution is Gaussian.
Since the number of degrees of freedom is
7 they correspond to $\Delta\chi^2=$8.2, 14.3 and 21.8
respectively. The parameters not shown in a given diagram are set to their best-fit value.

As one can see in Fig.1a  the iso-$\chi^2$ surface is rather prolate from
the low-$\Omega_m$ - high-$n$  corner to high-$\Omega_m$ - low-$n$.
This indicates a degeneracy in $n-\Omega_m$ parameter plane.  Within
the $1\si$ the 'maximum likelihood ridge' in this plane  can be
approximated by the equation $n\sqrt{\Omega_m}=0.73$. A similar  
degeneracy is observed in the $\Omega_{\nu}- \Omega_m$ plane
in the range $0\le\Omega_{\nu}\le 0.17$, $0.25\le\Omega_m \le 0.6$
(Fig.1c). The equation for the 'maximum likelihood ridge'
or 'degeneracy equation' has here the form:
$\Omega_{\nu}=0.023-0.44\Omega_m+1.3\Omega_m^2~$.

The important question is: which is the confidence limit of each parameter
marginalized over the other ones. The straightforward answer is the
integral of the likelihood function over the allowed range of all
the  other parameters.
But for a 6-dimensional parameter space this is computationally
time consuming.
Therefore, we have estimated the 1$\sigma$ confidence limits for
all parameters in
the following way. By variation of all parameter we determine
 the 6-dimensional $\chi^2$ surface which
contains  68.3\% of the total probability distribution.
We then project the surface onto  each axis
of parameter space. Its shadow on the parameter axes gives us the 1$\sigma$
confidence limits on cosmological parameters. For the best
$\Lambda$MDM model with one  sort of massive neutrinos the 1$\sigma$
confidence limits on parameters obtained  in this way are presented
in Table~\ref{tabmax}.
\begin{table}
\caption{\label{tabmax} The best fit values of all the parameters with
errors obtain by maximizing the (Gaussian) 68\% confidence contours
over all other parameters.}
\begin{center}
\begin{tabular}{||c|c||}
\hline
&\\
parameter & central value and errors\\ [4pt]
\hline
&\\
 $\Omega_m$ & $0.41^{+0.59}_{-0.22}$ \\[4pt]
$\Omega_{\nu}$ & $0.06^{+0.11}_{-0.06}$ \\[4pt]
 $\Omega_b$ & $0.039^{+0.09}_{-0.018}$ \\[4pt]
$h^{*)}$ & $0.70^{+0.15(+0.31)}_{-0.32}$ \\[4pt]
$n$ & $1.12^{+0.27}_{-0.30}$ \\[4pt]
 $b_{cl}$ & $2.22^{+1.3}_{-0.7}$\\[6pt]
\hline
\end{tabular}
\end{center}
 $ ^{*)}$ - the upper limit is obtained by including the
lower limit on the age of the Universe due to the age of oldest stars,
$t_0\ge13.2\pm 3.0$ \cite{car99}. The value obtained without this
constraint is given in parenthesis.
\end{table}

It must be noted that, using the observational data described above,
the upper $1\sigma$ edge for $h$ is equal 1.08 when we marginalized
over all other parameters. But this  contradicts
the age of the oldest globular clusters $t_0=13.2\pm 3.0$\cite{car99}. Thus
we have included this value into the marginalization procedure for
the upper limit of $h$.  We then have 8 degrees of freedom (24 data points)
and the 6-dimensional $\chi^2$ surface which contains
68.3\% of the probability is confined by the value 13.95.
We did not use the age of oldest
globular cluster  for searching of best fit parameters in general case because it is
only a lower limit for age of the Universe.

The errors given  in Table~\ref{tabmax}  represent 68\% likelihood, of course,
only when the probability distribution is Gaussian. As one can see
from  Fig.1 (all panels without degeneracy) the ellipticity of the likelihood
contours in most of planes is close to what is expected from a Gaussian
distribution. This indicates that around their maxima the likelihood
functions are close to Gaussian. However, the asymmetry of the error
bars obtained, shows that away from the maxima this is no longer the
case. Therefore, our estimates of the confidence limits have  to be
taken with a grain of salt.
The errors define the range of each parameter within which 
the best-fit values obtained for the remaining parameters lead to
$\chi^2_{min}\le 12.84$. Of course, the best-fit values of the 
remaining parameters lay within the range 
given in Table~\ref{tabmax}. However clearly not every 
set of parameters from these ranges satisfies  the condition,
$\chi^2_{min}\le 12.84$. For example,
standard CDM model ($\Omega_m=1$, $h=0.5$, $\Omega_b=0.05$, $n=1$ and best-fit value of
cluster biasing parameter $b_{cl}=2.17$ ($\sigma_8=1.2$)) has $\chi^2_{min}=142$ (!),
which excludes it at very high confidence level, $>99.999\%$. When we use the
baryon density inferred from nucleosynthesis 
($h^2\Omega_b=0.019$ ($b_{cl}=2.25$, $\sigma_8=1.14$))
the situation does not improve much, $\chi^2_{min}=112$. Furthermore,
even if  we leave $h$ as free parameter we still find $\chi^2_{min}=16$ 
($>1\sigma$) with the best-fit values  $h=0.37$ and $b_{cl}=3.28$ 
($\sigma_8=0.74$); this variant of CDM is ruled
out  by direct measurements of the Hubble constant.

The standard MDM model
($\Omega_m=1$, $h=0.5$, $\Omega_b=0.5$, $n=1$, $\Omega_{\nu}=0.2$, $N_{\nu}=1$
with a best value of the cluster biasing parameter 
$b_{cl}=2.74$ ($\sigma_8=0.83$)) does significantly better: it has 
$\chi^2_{min}=23.1$ ($99\%$ C.L.) which is out of the
$2\sigma$ confidence contour but inside $3\sigma$. With the
nucleosynthesis  constraint the situation does not change:
$\chi^2_{min}=22$; also if we leave $h$ as free parameter: 
$\chi^2_{min}=21$, $h=0.48$. But if, in addition, we let vary
$\Omega_{\nu}$, we obtain $\chi^2_{min}=13$
with best-fit values of $\Omega_{\nu}=0.09$, $h=0.43$, 
$b_{cl}=3.2$ ($\sigma_8=0.73$). This  means that the model is ruled out by the 
data set considered in this work at $\sim 70\%$ confidence level only. But also here the 
best-fit value for $h$ is very low. If we fix it at lower observational
limit $h=0.5$ then $\chi^2_{min}=18.9$ (the best fit values are: 
$\Omega_{\nu}=0.15$, $b_{cl}=2.8$ ($\sigma_8=0.83$)), which
corresponds to a confidence level of 95\% .

Therefore, we conclude
that the observational data set used here rules out  CDM models with 
$h\ge 0.5$, a scale invariant primordial
power spectrum ($n=1$) and $\Omega_k=\Omega_{\Lambda}=0$
at very high confidence level, $>99.99\%$. MDM models
with $h\ge 0.5$, $n=1$ and $\Omega_k=\Omega_{\Lambda}=0$
are ruled out at $\sim 95\%$ C.L.

One can see the 
model with one sort of massive neutrinos provides the best fit to 
the data, $\chi^2_{min}\approx 4.6$. Note, however, that there are 
only marginal differences in $\chi^2_{min}$ for $N_\nu = 
1,2,3$. Therefore, with the given accuracy of the data we cannot conclude 
whether -- if massive neutrinos are present 
at all -- their number of species is one, two, or three.  

The number of degrees of 
freedom is $N_F= N_{\rm exp}-N_{\rm par}= 7$. The
 $\chi^2_{min}$ for all cases is within the expected range,
$N_F-\sqrt{2N_F}\le \chi^2_{\min}\le N_F+\sqrt{2N_F}$ for the given
number of degrees of freedom. This means that the cosmological paradigm which
has been  assumed is consistent with the data.

One important question is how each point of the data influences our
result. To estimate this we have
excluded some data points from the searching procedure. 
Excluding any part of observable data results only in a  change
of the best-fit values of $n$, $\Omega_m$ and $h$ within the range of
their corresponding
standard errors. This indicates that the data are mutually in
agreement, implying consistent cosmological parameters (within the still
considerable error bars).
The small scale constraints, the Ly-$\alpha$ tests reduce the hot dark
matter content from
$\Omega_{\nu}\sim 0.22$ to $\sim 0.075$. The $\sigma_8$-tests further reduce
$\Omega_{\nu}$ to $\sim 0.06$. Including of the Abell-ACO power
spectrum in the search
procedure, tends to enhance  $\Omega_{\nu}$ slightly.
The most crucial test for the baryon content is of course the nucleosynthesis
constraint. Its $\sim 6\%-1\sigma$-accuracy safely keeps
$h^2\Omega_b$ near its median value 0.019. The parameter $\Om_b$ in
turn is only known to $\sim 36\%$ accuracy due to the large
errors of other experimental data used here, especially of the Hubble constant.
The accuracy of $h$ ($\sim17\%$) is better than the one
assumed from direct measurements, $\sim 23\%$.
 Summarizing, we conclude that all data points used here
are important for searching the best-fit cosmological parameters and do not contradict
each other.

Up to this point we ignored the uncertainties  in the COBE normalization.
Indeed, the statistical uncertainty of the fit to the
four-year COBE data, $\delta_h$, is 7\% (1$\sigma$) \cite{bun97} and
we want to study how this uncertainty
influences the accuracy of cosmological parameters which we determine?  

Varying $\delta_h$ in the 1$\sigma$ range we found that the best-fit values of all 
parameters except $\Omega_{\nu}$  do not vary by more than 2\% from the values presented
in Table 1. Only $\Omega_\nu$ varies in a range of 12\% . These 
uncertainties are significantly  smaller than the standard errors
given in Table 1 and neglecting them is thus justified.
The normalization constant
$A$, which for best model with one species of massive neutrinos, 
$A=4.68\cdot10^7(\Mpc)^{3+n}$, 
varies in a range of 8\%, and not 14\% as one might expect
 from (\ref{anorm}) at the first sight. 
The reason is that  variation of $\delta_h$ is somewhat compensated 
by correlated variation of $n$ and $h$.  Moreover, if we disregard the COBE 
normalization and treat the normalization constant $A$ as a free
parameter to be determined  like the others, its best-fit value
becomes $4.82 \cdot10^7(\Mpc)^{3+n}$ (for $N_{\nu}=1$), consistent
with COBE normalization. The best-fit values of the other parameters
correspondingly do not vary substantially: $n=1.09$, $\Omega_m=0.40$,
$\Omega_{\nu}=0.052$, $\Omega_b=0.041$, $h=0.68$ and $b_{cl}=2.19$
(this is less than 5\% except for $\Omega_{\nu}$, which is reduced by
$\sim 12\%$).  
This implies that determinations of the amplitude of scalar
fluctuations by the COBE measurement  of the large scale CMB
anisotropies  and by large scale structure data at much smaller scales
are in good agreement. It also indicates that a possible tensor
component in the COBE data cannot be very substantial.

\section{Conclusions}

We summarize, that the observational data of the LSS of the Universe
considered here can be
explained by a tilted $\Lambda$MDM inflationary model without tensor
mode. The best fit parameters are: 
$n=1.12\pm 0.09$, $\Omega_m=0.41\pm 0.11$, $\Omega_{\nu}=0.06\pm 0.028$,
$\Omega_b=0.039\pm 0.014$ and $h=0.70\pm 0.12$. 
All predictions of measurements are close to the experimental  
values given above and within the error bars of the data.  
The CDM density parameter is $\Omega_{cdm} = 0.31\pm0.12$ and
$\Omega_{\Lambda}$  is moderate, $\Omega_{\Lambda}=0.59\pm0.11$.
The neutrino matter density 
corresponds to a neutrino mass $m_{\nu}=94\Omega_{\nu}h^2\approx2.7\pm1.2$ eV.
The value of the Hubble constant is close to the measurements  by 
Madore et al.\cite{mad98}.
The age of the Universe for this model equals 12.3 Gyrs which is in
good  agreement with the age of the oldest objects in our galaxy \cite{car99}.
The spectral index coincides with the 
COBE prediction. The relation between the matter density 
$\Omega_m$ and the cosmological
constant $\Omega_{\Lambda}$ agrees well with the independent 
measurements of cosmic 
deceleration and global curvature based on the SNIa observation.

The 1$\sigma$ (68.3\%) confidence limits on each cosmological
parameter, obtained by marginalizing over the other parameters, are
 $0.82\le n\le 1.39$, $0.19\le\Omega_m\le 1$, $0\le\Omega_{\Lambda}\le0.81$,
$0\le\Omega_{\nu}\le 0.17$, $0.021\le\Omega_b\le 0.13$ and
$0.38\le h\le 0.85$.

The observational data set used here rules out the CDM models with $h\ge 0.5$, 
scale invariant primordial power spectrum $n=1$ and $\Omega_{\Lambda}=\Omega_k=0$
at very high confidence level, $>99.99\%$. Also pure MDM models are ruled out 
at $\sim 95\%$ C.L. 

It is remarkable also that this  data  set
determines the value of normalization constant for scalar
fluctuations which approximately equals the value deduced from
COBE four-year data. It  indicates that a possible tensor component
in the COBE data cannot be very substantial.  

The coincidence of the values of cosmological parameters
obtained by different methods indicates that a wide set of 
cosmological measurements are correct and that their theoretical 
interpretation is consistent. However, we must also 
note that the accuracy of present
observational data on the large scale structure of the Universe is
still insufficient to determine a set of cosmological parameters with
high accuracy.

\end{document}